# APPLICATION OF FRACTIONAL DERIVATIVE OPERATORS TO ANOMALOUS DIFFUSION and PROPAGATION PROBLEMS

## Andrzej J. Turski, Barbara Atamaniuk and Ewa Turska

Institute of Fundamental Technological Research, PAS, Świętokrzyska 21, PL- 00-049 Warsaw, Poland, aturski@ippt.gov.pl

+

***Abstract*** *-* We investigate evolution equations for anomalous diffusion employing fractional derivatives in space and time. Linkage between the space-time variables leads to a new type of fractional derivative operator. Fractional diffusion equations account for typical "anomalous" features which are observed in many systems, e.g. in the case of dispersive transport in amorphous semiconductors, liquid crystals, polymers, proteins and biosystems. In contrast to Gaussian diffusion, fractional diffusion is related to LÉVY STABLE NON-GAUSSIAN PROCESSES. The typical features of the processes are heavy tails of probability density distributions. Conservation laws in relation to fractional operators are discussed.

The next objective of this paper is an application of main rules of fractional calculus, fractional Laplacians, factorization of the Helmholtz equation to obtain four pairs of fractional eigenfunctions allowing to construct a solution to the well known half-plane diffraction problems. Factorizing the Leontovich-Fock equation, (parabolic wave equation-PWE), we determine semi-differential fractional solutions, which allow us to find paraxial solutions for given beam boundary conditions*.*

# INTRODUCTION

Integration and differentiation to an arbitrary order named fractional calculus has a long history, see [1, 2, 11]. The concept of non-integral order of integration can be traced back to the genesis of differential calculus itself: the philosopher and creator of modern calculus, the Newton's rival Leibniz made some remarks on the meaning and possibility of fractional derivative of order ½ in the late of 17-th century. However, a rigorous investigation was first carried out by Liouville in a serious of papers from 1832-1837, where he defined the first outcast of an operator of fractional integration.

In the background of the fractional operations, we see generalization of integer order calculus to an arbitrary order, class of fractionally differentiable functions (called "differintegrable functions" see, [1, 2, 11], where use of the term is widely discussed) and applications of the calculus.

The mathematical theory of the fractional calculus and the theory of fractional ODE are well developed and there is a vast literature on the subject, e.g. see; [1], [2], [3], [4], and [11]. The theory of PDE equations is a recently investigated problem and the theory mainly concerns fractional SUPER- and SUB-DIFFUSION equations, e.g. see; [5], [6], [7], [8] and [9].

# FRACTIONAL DIFFUSION

In classical diffusion, particles spread in a normal bell-shaped pattern according to the Gaussian probability distribution. Anomalous diffusion occurs when the growth rate or shape of the particle distribution is different than Gaussian.

Anomalous diffusion is observed in many physical situations, motivating the development of new mathematical and physical models. Fractional diffusion equations account for typical "anomalous" features which are observed in many systems, e.g. in the case of dispersive transport in amorphous semiconductors, glasses, liquid crystals, polymers, biopolymers, proteins, biosystems and even in ecosystems.

1. Very large particle jumps are associated with fractional derivatives in space.
2. Very long waiting times lead to fractional derivatives in time.
3. Empirical evidence shows that the waiting time between jumps can be correlated with the ensuing size of the particle jumps. In the continuous time random walk, the size of particle jumps can depend on waiting time between jumps. For these models, the limiting particle distribution is governed by a fractional differential equation involving *coupled space-time fractional derivative operators.*

Let $W(x,t)$ denote the propagator or the relative concentration of particles at location $x \in R$ at time $t$. The classical diffusion (transport) equation is of the following form:

$$\partial_t W(x,t) = \frac{1}{2} \partial_x^2 W(x,t) \qquad (1)$$

The equation can be solved using the Fourier transform:

$$w(k,t) = \int_{-\infty}^{\infty} e^{ikx} W(x,t) dx.$$

This converts the diffusion equation to an ordinary differential equation

$$\frac{dw}{dt} = \frac{1}{2}(-ik)^2 w(k,t).-$$

The initial condition $w(k,0) \equiv 1$ is equivalent to $W(x,0) = \delta(x)$, so that all particles start at position $x = 0$ at time $t = 0$. The solution:

$$w(k,t) = \exp(-ik^2 t)$$

is a characteristic function of the normal probability density distribution:

$$W(x,t) = (2\pi t)^{-1/2} e^{-x^2/2t},$$

with mean zero and standard deviation $\sqrt{t}$. This is also, using Central Limit Theorem (CLT), the limiting density of a random walk of particle jumps, where the jumps have mean zero and variance one, and hence the standard Brownian motion model emerged.

The probability distribution function (pdf) of the particle jumps has symmetric regularly varying tails with index $-\alpha$ for some $0 < \alpha < 2$. This means that the probability of jumping a distance greater than $r$ falls off like $r^{-\alpha}$. The classical CLT does not apply when the probability distribution functions (pdf) are non-Gaussian.

Let explain the behavior of stable probability functions (Levy's distributions) by the following figures:

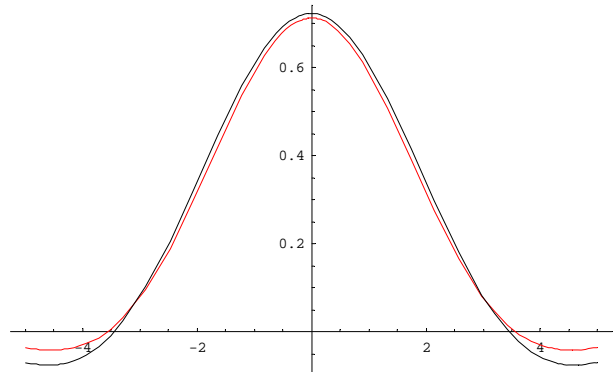

Figure 1. α=3 (red), 4 (black).

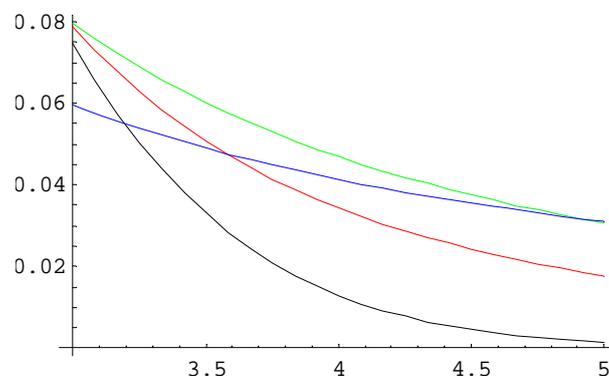

Figure 2. Pareto (heavy) tails: α=0.5 (blue), 1 (green), 1.5 (red), 2(black).

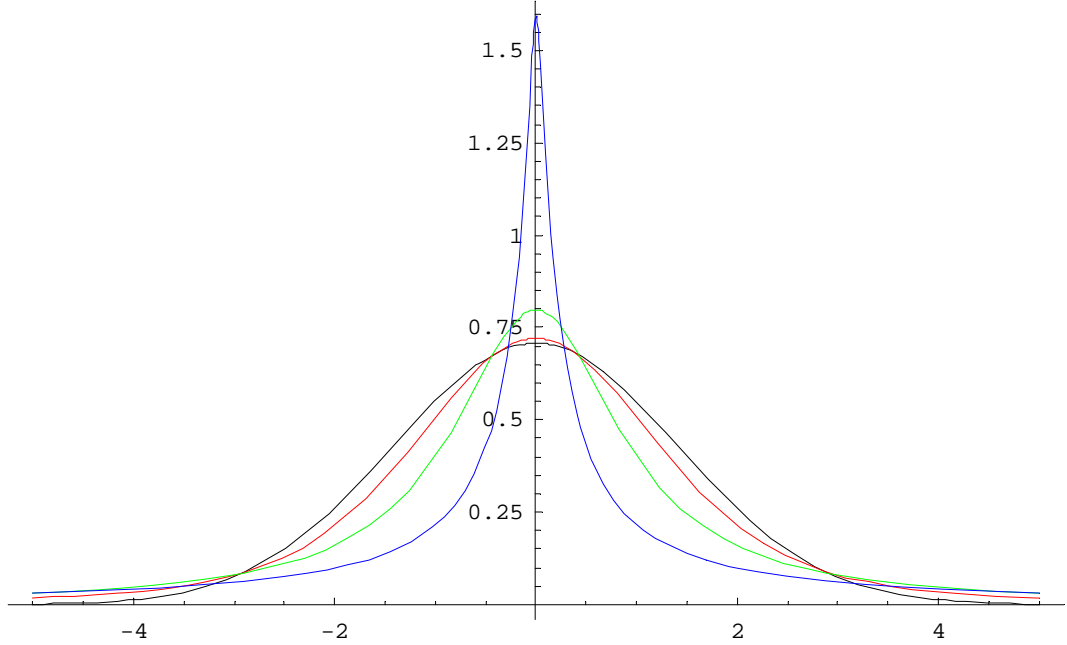

Figure 3. α=0.5 (blue), 1 (green), 1.5 (red), 2 (black).

A generalized CLT implies that the random walk converges to a stable Levy motion whose probability density $W(x,t)$ has a characteristic function (Fourier transform) in the following form:

$$w(k,t) = \exp(-|k|^\alpha t),$$

which is evidently the solution to the equation:

$$\frac{dw(k,t)}{dt} = -|k|^\alpha w(k,t) \text{ with } w(k,0) \equiv 1.$$

Inverting shows that the particle concentration (propagator) solves a fractional partial differential equation:

$$\partial_t W(x,t) = \partial_{|x|}^\alpha W(x,t).$$

The symmetric fractional derivative operator $\partial_{|x|}^\alpha$ corresponds to multiplication by the symbol $-|k|^\alpha$ in the Fourier space. This is usual fractional power of the second order derivative operator. Asymmetric particle jumps lead to a more general form of fractional derivative operator:

$$p\partial_x^\alpha + q\partial_{-x}^\alpha$$

with the symbol $p(-ik)^\alpha + q(ik)^\alpha$ in the Fourier space, where $p = 1-q$ is the asymptotic fraction of positive jumps as the jump tends to infinity. For symmetric vector jumps a similar arguments lead to the equation containing fractional Laplacian:

$$\partial_t W(\vec{x},t) = \Delta_x^{\alpha/2} W(\vec{x},t)$$

with the symbol $-\|k\|^\alpha$ in the Fourier space.

The following symmetric α-stable (S$\alpha$S) distributions of time series are depicted in the figure 4 for comparison purposes. One can observe Gaussian time series, closed to Gaussian time series and the time series with long waiting time and strong jumps. The last processes are heavy tailed and there is none variance. As $\alpha$ decreases both the occurrence rate and the strengths of the outliers increases, resulting to very impulsive processes. S$\alpha$S time series obey two important properties: stability and generalized CLT. The properties justify the role of the processes in data modeling.

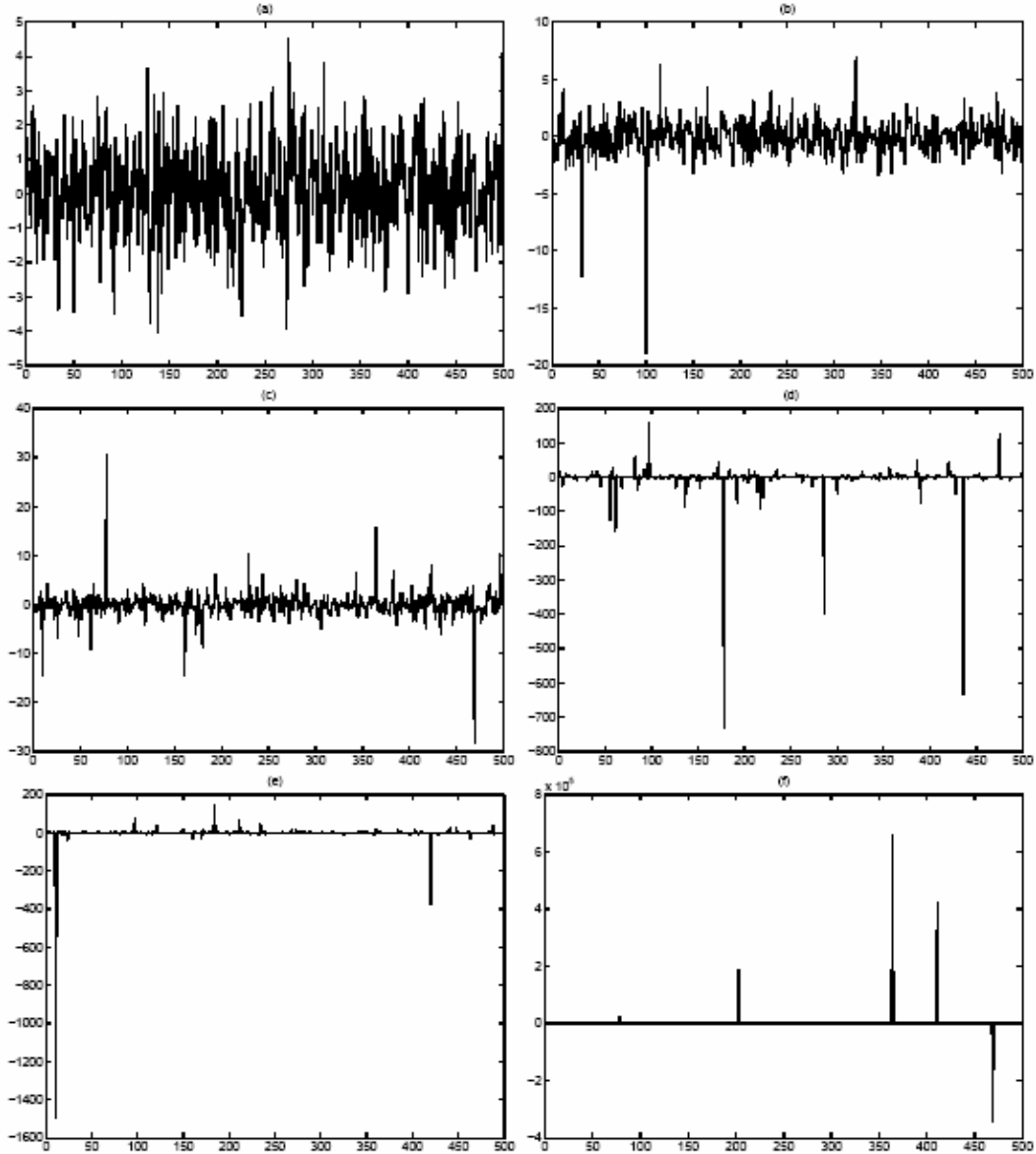

Figure 4. $S\alpha S$ time series. (a): α=2.0 (Gaussian), (b): α=1.95, (c): α=1.5, (d): α=1.0, (e): α=0.85, (f): α=0.45. (see: [12])

If the time distribution between particle jumps is heavy tailed and varies regularly with the index $0 < \beta < 1$ (roughly speaking, the chance of waiting longer than time $t$ before the next jump falls off like $t^{-\beta}$) then the random walk of particle jumps (distributed as before) converges to a Lévy motion.

Assuming that the waiting time and the ensuing particle jump are independent, and the governing equation becomes:

$$\partial_t^\beta W(x,t) = \partial_{|x|}^\alpha W(x,t) + \frac{W(x,0)t^{-\beta}}{\Gamma(1-\beta)}.$$

The equation was first proposed by Zaslavsky [10] as a model for Hamiltonian chaos. Asymmetric jumps, or vector jumps, modify the spatial derivative in the same manner as before. Heavy tailed particle jumps lead to fractional derivatives in space, and heavy tailed waiting times introduce fractional derivatives in time.

When the waiting times and the particle jumps are dependent random variables, a different form of the governing equation emerges. The limiting process is still a Lévy motion subordinated to an inverse stable subordinator, but now the two processes are dependent. The space-time vector consisting of the waiting time and the jump has to be treated in a proper way, since each coordinate has a different tail behavior [3]. This leads to a governing equation that employs a new kind of coupled space-time fractional derivative. Suppose the waiting time $J$ satisfies $P(J > t) = t^{-\beta}$ for large $t$ and the symmetric particle jump size $Y$ is normally distributed with mean zero and variance $2t$ when the waiting time $J = t$. Then the governing equation:

$$(\partial_t + \partial_x^2)^\beta W(x,t) = \frac{W(x,0)t^{-\beta}}{\Gamma(1-\beta)}$$

employs a coupled space-time fractional derivative with Fourier-Laplace symbol $(s + k^2)^\beta$. What makes this problem interesting is that, since space and time are inexorably linked, one cannot view these evolution equations in the usual manner, as PDE on some abstract function space.

This Chapter needs some deep insight comments. The most important properties of physical systems are conservation laws, like energy, charge, momentum and angular momentum conservations. The Neother's theorem is a central result in theoretical physics that express the one-to-one correspondence between *symmetries* and c*onservations laws*. This exact equivalence holds for all physical laws, (e.g., integer order ODE, PDE and I.E.) based upon *action principles* defined over a *symplectic space*. It is named after the early 20-th century mathematician Emmy Noether.

In the case of fractional derivative operators describing physical events, we have to do at most with semi-groups. There is no proper *"symmetry"* notion, neither *"action principles"* (e.g., Hamiltonian's principle) nor *symplectic* topology and since there is no notion of conservation laws. It means that we have to deal with a different physics.

There is some resemblance between fractional operators and irrational/transcendental numbers. The law of order in application to rational numbers is simple and can be realized by use of ordinal numbers. But in the case of ordering irrational numbers we need some rules like *"Dedekind's crossections"* or *"Cantor's methods"* based upon *Cauchy sequences*. We note that analogous rules are expected to be elaborated in relation to the factional derivative operators, which are connected to systems of non-Gaussian statistics. There is no energy and higher moments in the systems and their conservations are meaningless. The statistical approach delivers two important factors: stable evolution of the processes and generalized CLT.

## **FACTORIZATION OF HELMHOLTZ AND PW EQUATIONS**

An eigenfunction $y(\xi)$ of the linear operator $L[y(\xi)]$ is such a function that the repeated operations preserve the function, e.g. $L[y(\xi)] = Cy$ with the exactness to a multiplicative constant $C$. In the case of fractional operation, the definition is extended to preservation of the function but an additive constant or a term of power of $\xi$, e.g. $(\pi\xi)^{-1/2}$, is subtracted at each step, see [4].

Factorizing the 1-D Helmholtz equation:
$$(_0D_x^2 + k^2)y(x) = 0, \qquad (2)$$

where $_0D_x^2$ is the "fractional" derivative operator on the (0,1] interval. Let factorize the operator:

$$_0D_x^2 + k^2 \equiv (_0D_x^1 + ik)(_0D_x^1 - ik) \equiv$$
$$(_0D_x^{1/2} + \sqrt{ik})(_0D_x^{1/2} - \sqrt{ik})(_0D_x^{1/2} + \sqrt{-ik})(_0D_x^{1/2} - \sqrt{-ik}) \qquad (3)$$

and in the Laplace symbol space $s$, we have:

$$_0D_x^\alpha \Rightarrow s^\alpha \quad and \quad s^{1/2} + \sqrt{ik}.$$

The solution to the fractional : $(_0D_x^{1/2} \pm \sqrt{ik})y(x) = 0$, takes the form:

$$y(x) = \frac{1-i}{\sqrt{2\pi kx}} + \frac{2}{\sqrt{\pi}} e^{ikx} \int_{\sqrt{kx}}^{\infty} e(\pm it^2) dt.$$

In the sense of the eigenfunction definition given above, and rearranging the complex and complex conjugate terms, we can write the following real modes:

$$\sin kx, \quad \cos kx, \quad \int_{\sqrt{kx}}^{\infty} \cos(kx - t^2) dt, \quad and \quad \int_{\sqrt{kx}}^{\infty} \sin(kx - t^2) dt.$$

In the case of the 2-D Helmholtz equation:
$$(\Delta_{x,y} + k^2)\Phi(x,y) = 0, \qquad (4)$$

we denote $|\vec{k}| = \sqrt{a^2 + b^2}$ and $|\vec{r}| = \sqrt{x^2 + y^2}$. In the half-plane case *(half-plane diffraction)* for the 2-D Laplace symbol space, we have:

$$\Delta_{x,y}^{\alpha/2} \Rightarrow (s_x^2 + s_y^2)^{\alpha/2} \equiv \|s\|^{\alpha/2}, \quad 0 < \alpha \le 2 \qquad (5)$$

Processing like in the 1-D case, we derive the full set of real modes of the Helmholtz equation:

$$\sin(ax \pm by), \quad \cos(ax \pm by)$$

$$\int_{u(x,y)}^{\infty} \sin(ax \pm by - t^2)dt \quad \text{and} \quad \int_{u(x,y)}^{\infty} \cos(ax \pm by - t^2)dt,$$

We note that $u(x,y) = \sqrt{\|\vec{k}\|\|\vec{r}\| + ax \pm by}$.

The fractional eigenfunctions represent "edge waves", which are related to half-plane and can be used to construct the solutions of diffraction problems. It deserves comment, that the first two modes are derived by a separation of variables $(x, y)$ and then by the first factorization. The second two modes are derived by the second factorization and are really 2-D modes since their variables $(x, y)$ are non-separable. We note that for 3-D Helmholtz equation such solution is not known. The known solution can be reduced to the 2-D problem, see [4].

If the wave equation:

$$(\Delta_{\vec{x}} - \frac{1}{c^2}\partial_t^2)\Psi(\vec{x},t) = 0 \qquad (6)$$

is Fourier transform in time and Laplace transform in the half-space, then the Laplace-Fourier space symbol is:

$$\|s\|^\alpha - \|\vec{k}\|^\alpha,$$

where $\|\vec{k}\| = \sqrt{\sum_j^n k_j^2} = \frac{\omega}{c}$, $\|s\| = \sqrt{\sum_j^n s_j^2}$ and $|\vec{x}| = \sqrt{\sum_j^n x_j^2}$.

Factorizing twice Eq. (6), like in the case of Helmholtz equations, the following fractional solution to the equation is derived:

$$\Psi(\vec{r},t) = \int_A^{u(\vec{x},t)} \exp(it^2)dt,$$

where $u(\vec{x},t) = \sqrt{\|\vec{k}\|ct \pm \vec{k}\cdot\vec{x}}$ and $A$ as well as $k_j$ are arbitrary real constants. The number of $k_j$ constants is equal to the number of space dimensions. The solution can not be derived by a separation of variables and the first factorization but it is a particular solution of the well known general solution to the Eq. (6).

Considering the two cases of unidirectional $(z > 0)$ propagation and transversal beam dimensions commonly denoted as: $(1+1)D$ and $(1+2)D$, we derived the following equations:

$$(2ik\partial_z^1 + \Delta_{x,y})u(x,y,z) = 0 \qquad (7)$$

and

$$(2ik\partial_z^1 + {}_{-\infty}D_x^2)u(x,z) = 0. \qquad (8)$$

Next, we factorize the Eq. (8) to obtain:

$$_{-\infty}D_x^2 + 2ik_0 D_z^1 = [_{-\infty}D_x^1 + \sqrt{2k}e^{-i\pi/4}{}_0 D_z^{1/2})][_{-\infty}D_x^1 - \sqrt{2k}e^{-i\pi/4}{}_0 D_z^{1/2}],$$

where the fractional derivative, according to the Riemann-Liouville definition, takes the form:

$$_0 D_x^{1/2} u(x,z) = \frac{1}{\sqrt{\pi}} \partial_z \int_0^z u(x,\varsigma) \frac{d\varsigma}{\sqrt{z-\varsigma}}.$$

The fractional differential operator:

$$_{-\infty}D_x^1 \pm \sqrt{2k} \exp(-i\pi/4){}_0 D_z^{1/2} \equiv L^{1/2},$$

employs a coupled fractional derivative with Fourier-Laplace transform symbol: $\kappa \pm (2k)^{1/2} s^{1/2} \exp(i\pi/4)$, where $\kappa$ is responsible for Fourier transform with respect to the $x$-axis, s is related to Laplace transform in $z$ and $k$ is a constant of propagation in the z- direction. The solution to the fractional equation:

$$L^{1/2} u(x,z) = 0,$$

takes the form:

$$u(x,z) = \frac{x\sqrt{k}}{\sqrt{2\pi i z^3}} \exp(\frac{ikx^2}{2z}).$$

This is well-known solution to the PWF equation (8). The second factorization of the equation leads to following fractional new solution:

$$u(x,z) = \frac{1}{k}\sqrt{\frac{i}{\pi}} \int_{x\sqrt{\frac{k}{2z}}}^{\infty} \exp(it^2) dt \qquad (9)$$

Factorizing twice $(1+2)D$ equation (7), we derive the Fourier-Laplace symbol:
$\|\kappa\|^{1/2} \pm (2k)^{1/4} s^{1/4} \exp(-i\pi/8)$, where $\|\kappa\| = \sqrt{\kappa_x^2 + \kappa_y^2}$.

Hence, we derive following fractional solutions:

$$G(x,y,z) = \frac{i}{\pi} (\int_{x\sqrt{\frac{k}{2z}}}^{\infty} e^{it^2} dt)(\int_{y\sqrt{\frac{k}{2z}}}^{\infty} e^{it^2} dt) \quad \text{or}$$

$$G(x,y,z) = \frac{k}{\pi} \int_{\frac{x+y}{2}\sqrt{\frac{k}{z}}}^{\infty} e^{it^2} dt.$$

The derivatives $\partial_x^m \partial_y^n$ of G(x, y, z), where *m* and *n* are natural numbers, satisfy the PWE and may be related to higher order Gaussian-Hermite optical beams. By use of the fractional functions, we may obtain known and new solutions for optical beams.

## CONCLUSION

- Anomalous diffusion and advection are observed in many physical situations, motivating the development of new mathematical and physical models. Fractional diffusion equations account for typical "anomalous" features which are observed in many systems, e.g. in the case of dispersive transport in amorphous semiconductors, glasses, liquid crystals, polymers, biopolymers, proteins, biosystems- and even in ecosystems.
- Large particle jumps and waiting times of particles lead to stable probability functions (Levy's pdf) with heavy tails and undefined variance.
- When the waiting times and the particle jumps are dependent random variables, a coupled space time fractional derivative operators emerge.
- The second factorization of the 2-D HELMHOLTZ equations leads to really 2-D solutions, which represent the "edge waves".
- The second factorization of the (2+1)-D Parabolic Wave Equations (Schrödinger type) leads to "mother solutions". They allow deriving all known solutions by a proper differentiation with respect to the variables, which are transverse to the axis of propagation.
- Here considered fractional derivative operators have Fourier-Laplace symbols.